\documentclass[12pt,a4paper]{article}

\usepackage{tikz}
\usepackage{authblk}
\usepackage{amsmath,amssymb}
\usepackage{graphicx}
\usepackage{amsfonts}
\usepackage{epsfig}
\usepackage{times,pstricks,pst-node}
\usepackage{pst-all} 
\usepackage{tikz}
\usepackage{pgflibrarysnakes}
\usetikzlibrary[snakes]

\newcommand{\del}{\partial}

\newcommand{\wt}{\widetilde}
\newcommand{\mc}{\mathcal}
\newcommand{\ms}{\mathsf}
\newcommand{\dd}{\mathrm{d}}

\newcommand{\ket}[1]{\left|#1\right\rangle}      % Ket-Zustand
\newcommand{\bra}[1]{\left\langle #1\right|}     % Bra-Zustand
\newcommand{\eq}{\begin{equation}}
\newcommand{\en}{\end{equation}}
\newcommand{\bear}{\begin{eqnarray}}
\newcommand{\ear}{\end{eqnarray}}

\usetikzlibrary{arrows,shapes,positioning}
\usetikzlibrary{decorations.markings}
\tikzstyle arrowstyle=[scale=1]
\tikzstyle directed=[postaction={decorate,decoration={markings,
		mark=at position .65 with {\arrow[arrowstyle]{stealth}}}}]
\tikzstyle reverse directed=[postaction={decorate,decoration={markings,
		mark=at position .65 with {\arrowreversed[arrowstyle]{stealth};}}}]

\title{Boundary correlations for the six-vertex model with reflecting end boundary condition}

\author{I.R. Passos and G.A.P. Ribeiro\footnote{E-mail: pavan@df.ufscar.br}}
\affil{Departamento de F\'{i}sica, Universidade Federal de S\~ao Carlos \\ S\~ao Carlos, SP 13565-905, Brazil}
\date{}
  
\begin{document}

\maketitle
\thispagestyle{empty}

\begin{abstract}
We consider the six-vertex model with reflecting end boundary condition. We compute 
analytically boundary correlation functions, such as the boundary polarization and the emptiness formation probability. In order to do that, we use the Sklyanin's reflection algebra to derive recursion relations for the partition function of the model as well as for the boundary correlations in terms of the partition function. Thanks to the Tsuchiya determinant formula, these recursion relations allow the boundary correlations to be also efficiently written in determinant form.
\end{abstract}

\newpage

\pagestyle{plain}
\pagenumbering{arabic}

\section{Introduction}

The six-vertex model is one of the most important models in the realm of integrable systems \cite{BAXTER,BOOK}. Originally proposed to investigate the residual entropy of the ice \cite{PAULING}, the versatility of the model allowed its use
in different contexts, ranging from classical statistical mechanics \cite{BAXTER} to applied mathematics \cite{KUPERBERG}. Moreover, the ice-rule, which characterizes the model, has been observed in frustrated magnetic systems called spin ices \cite{BRAMWELL}. It is worth to mention that recent developments involving artificial spin systems have also 
fostered the pursuit of realization of vertex systems through the experimental approach devised in \cite{WANG}. 

From the theoretical point of view, the six-vertex model \cite{BAXTER,BOOK} and the dependency of its physical properties on boundary conditions have been largely studied over the years. This model was investigated under periodic, anti-periodic and a number of fixed boundary conditions \cite{LIEB,WU,OWCZAREK,BATCHELOR,KOREPIN2000,ZINNJUSTIN,BLEHER,RIBEIRO2015b,PARTIAL,GALLEAS,BLEHER2017}. 

The first clear instance of boundary condition which confirmed such special feature of the six-vertex model, in the thermodynamic limit, is the domain wall boundary condition \cite{KOREPIN1982,KOREPIN1992}. It was proven that the thermodynamic properties like free energy for domain wall boundary condition do differ from the result for periodic boundary conditions \cite{KOREPIN2000,ZINNJUSTIN,BLEHER}. This fact suggested the existence of spatial phase separation, which was investigated numerically \cite{NUM-DWBC}. 

Because of the ice rule, the number of configurations in the system can be considerably restricted for 
fixed boundary conditions \cite{KOREPIN2000,RIBEIRO2015b}. In this case, even when the parameters of the system are adjusted for the disordered regime, the boundary conditions may induce the formation of ordered regions of macroscopic size spreading towards the bulk of the lattice. In the thermodynamic limit, the curves that separate the ordered and disordered regions are called arctic curves. For the six-vertex model with domain wall boundary condition, analytical expressions for the arctic curves in the disordered regime were obtained \cite{ARTIC}.

In order to analytically derive the arctic curves for the six-vertex model with domain wall boundary condition a lot of machinery is needed. For instance, the determinant representation for the partition function \cite{KOREPIN1992}, boundary correlations \cite{BOGOLIUBOV,PRONKO} and the emptiness formation probability \cite{COLOMO} were essential to characterize the spatial separation line between the ferroelectric and disordered regions. 

There is another instance of boundary with the potential for analytical results, the so-called reflecting end boundary condition \cite{TSUCHIYA}. In this case, one can build the partition function of the six-vertex model 
from the Bethe state defined in Sklyanin construction for open spin chains \cite{SKYLIANIN}. The partition function with reflecting end boundary can also be represented as a determinant \cite{TSUCHIYA}. This allowed for the calculation of the free energy in the thermodynamic limit \cite{RIBEIRO2015a}, which again differs from the case of periodic boundary condition.

Nevertheless, there are still no results for boundary correlations and emptiness formation probability for the case of reflecting end boundary. Besides of its own relevance, the knowledge of boundary correlation functions could be useful in the analytical investigation of the arctic curves. It is worth noting that the spatial phase separation for the reflecting end boundary was recently studied numerically \cite{LYBERG} and the results confirm the expectations for phase separation, however, there are no analytical results for the arctic curves. 
Although  it seems not possible to have full control on the boundary conditions at the current experimental level, the reflecting end boundary condition is an important case to study, since its determinant form \cite{TSUCHIYA} allows for exact analytical results for the physical quantities at finite lattice sizes and also in the thermodynamic limit. 

In this paper, we present the analytical derivation of boundary correlation functions and the emptiness formation probability for the six-vertex model with reflecting end boundary on a $2 N \times N$ lattice.
These correlations are represented as $N\times N$ determinants
and their homogeneous limit is taken. 
Thanks to the machinery of biorthogonal polynomials, such determinant representations can be further reduced. In particular, for the emptiness formation probability corresponding to a $2(N-r)\times s$ frozen region, its expression is rewritten in terms of the determinant of a $s\times s$ matrix.

This paper is organized as follows. In section \ref{sixvertex}, we introduce the six-vertex model and the reflecting end boundary. In section \ref{boundcorr}, we discuss the Sklyanin reflection algebra and we compute two different kinds of boundary correlation functions, whose homogeneous limit is taken. The evaluation of the emptiness formation probability is presented in section \ref{EFP}. Finally, our
conclusions are given in section \ref{conclusion}.

\section{The six-vertex model and the reflecting end boundary condition}\label{sixvertex}

In this section we describe the six-vertex model with reflecting end boundary conditions. 

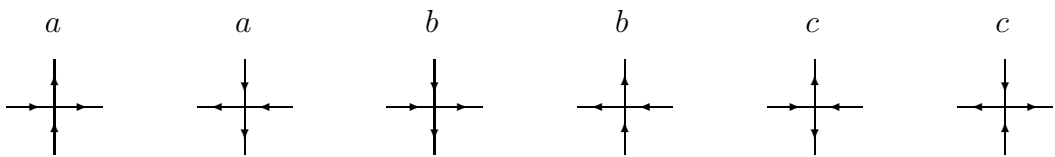
\begin{figure}[h]
	\unitlength=1.25mm
	\begin{center}
		\begin{picture}(120,20)
		\put(4,13){$a$}
		\put(5,0){\line(0,1){10}}
		\put(5,7.5){\vector(0,1){1}}
		\put(5,2.5){\vector(0,1){1}}
		\put(0,5){\line(1,0){10}}
		\put(2.5,5){\vector(1,0){1}}
		\put(7.5,5){\vector(1,0){1}}
		\put(24,13){$a$}
		\put(25,0){\line(0,1){10}}
		\put(25,7.5){\vector(0,-1){1}}
		\put(25,2.5){\vector(0,-1){1}}
		\put(20,5){\line(1,0){10}}
		\put(22.5,5){\vector(-1,0){1}}
		\put(27.5,5){\vector(-1,0){1}}
		\put(44,13){$b$}
		\put(45,0){\line(0,1){10}}
		\put(45,2.5){\vector(0,-1){1}}
		\put(45,7.5){\vector(0,-1){1}}
		\put(40,5){\line(1,0){10}}
		\put(42.5,5){\vector(1,0){1}}
		\put(47.5,5){\vector(1,0){1}}
		\put(64,13){$b$}
		\put(65,0){\line(0,1){10}}
		\put(65,2.5){\vector(0,1){1}}
		\put(65,7.5){\vector(0,1){1}}
		\put(60,5){\line(1,0){10}}
		\put(62.5,5){\vector(-1,0){1}}
		\put(67.5,5){\vector(-1,0){1}}

		\put(84,13){$c$}
		\put(85,0){\line(0,1){10}}
		\put(85,7.5){\vector(0,1){1}}
		\put(85,2.5){\vector(0,-1){1}}
		\put(80,5){\line(1,0){10}}
		\put(82.5,5){\vector(1,0){1}}
		\put(87.5,5){\vector(-1,0){1}}
		\put(104,13){$c$}
		\put(105,0){\line(0,1){10}}
		\put(105,7.5){\vector(0,-1){1}}
		\put(105,2.5){\vector(0,1){1}}
		\put(100,5){\line(1,0){10}}
		\put(102.5,5){\vector(-1,0){1}}
		\put(107.5,5){\vector(1,0){1}}
		\end{picture}
		\caption{The Boltzmann weights of the six-vertex model.}
		\label{6vert}
	\end{center}
\end{figure}

The six-vertex model is a classical vertex model in a rectangular lattice \cite{BAXTER,BOOK}, whose six allowed configurations are depicted in Figure \ref{6vert}.

In the realm of integrability, the Boltzmann weights $a(\lambda), b(\lambda)$ and $c(\lambda)$ can be seen as matrix elements of the $R$-matrix,
\eq
R(\lambda)=\left(\begin{array}{cccc}
	a(\lambda) & 0 & 0 & 0 \\
	0 & b(\lambda)& c(\lambda) & 0  \\
	0 & c(\lambda) & b(\lambda) & 0 \\
	0 & 0 & 0 & a(\lambda)
\end{array} \right),
\en
which is, in turn, a solution of the Yang-Baxter equation, 
\eq
R_{12}(\lambda_1-\lambda_2)R_{13}(\lambda_1) R_{23}(\lambda_2) =R_{23}(\lambda_2)R_{13}(\lambda_1) R_{12}(\lambda_1-\lambda_2).
\label{yangbaxter}
\en
It is worth to note that the Yang-Baxter equation constraints the Boltzmann weights such that,
\eq
\Delta=\frac{a^2+b^2-c^2}{2 a b},
\label{inv}
\en
where $\Delta$ is constant.

The ordered product of the Boltzmann weights along the $j$-th row of the rectangular lattice gives rise to the monodromy matrix ${\cal T}_{j}(\lambda)=R_{j N}(\lambda-\mu_N)\cdots R_{j 1}(\lambda-\mu_1)$,
which can be seen as a $2\times 2$ matrix on the horizontal space also known as auxiliary space. It can be represented as
\eq
{\cal T}(\lambda)=\left(\begin{array}{cc}
	A(\lambda) & B(\lambda) \\
	C(\lambda) & D(\lambda) 
\end{array}\right),
\en
where the matrix elements $A(\lambda)=A(\lambda,\{\mu\})$, $B(\lambda)=B(\lambda,\{\mu\})$, $C(\lambda)=C(\lambda,\{\mu\})$ and $D(\lambda)=D(\lambda,\{\mu\})$ are operators acting non-trivially on the vertical space also referred as quantum space. The above monodromy matrix elements play an important role in the Yang-Baxter algebra, which is used to diagonalize the transfer matrix of the six-vertex model with periodic boundary condition and consequently to deal with the partition function of the six-vertex model with periodic boundary condition along horizontal and vertical direction.

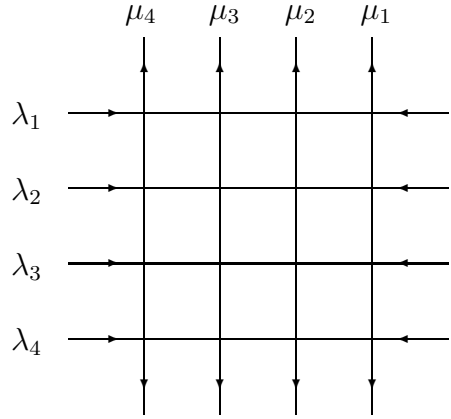
\begin{figure}[h]
	\unitlength=0.5mm
	\begin{center}
		\begin{picture}(100,100)(-30,-10)
		
		\multiput(-20,0)(0,20){4}{\line(1,0){100}}
		\multiput(0,-20)(20,0){4}{\line(0,1){100}}
		% arrow
		\multiput(-7.5,0)(0,20){4}{\vector(1,0){1}}
		\multiput(67.5,0)(0,20){4}{\vector(-1,0){1}}
		\multiput(0,-12.5)(20,0){4}{\vector(0,-1){1}}
		\multiput(0,72.5)(20,0){4}{\vector(0,1){1}}
		
		\put(-35,-3){$\lambda_4$}
		\put(-35,17){$\lambda_3$}
		\put(-35,37){$\lambda_2$}
		\put(-35,57){$\lambda_1$}
		
		\put(-5,85){$\mu_4$}
		\put(17,85){$\mu_3$}
		\put(37,85){$\mu_2$}
		\put(57,85){$\mu_1$}
		
		\end{picture}
	\end{center}
	\caption{$Z_N^{DWBC}$: the partition function for $N=4$ of the six-vertex model with domain wall boundary condition.}
	\label{DWBC}
\end{figure}

On the other hand, the partition function of the six-vertex model in a square lattice $N\times N$ with domain wall boundary condition is defined, e.g. in terms of products of the monodromy matrix element $B(\lambda_j)$ as follows (see Figure \ref{DWBC}),
\eq
Z_N^{DWBC}(\{\lambda\},\{\mu\})=\bra{\Downarrow}B(\lambda_N)\cdots B(\lambda_2) B(\lambda_1)\ket{\Uparrow},
\en
where $\ket{\Uparrow}=\ket{\uparrow\cdots\uparrow}$ and $\ket{\Downarrow}=\ket{\downarrow\cdots\downarrow}$ are the up and down ferromagnetic states.

\begin{figure}[h]
	\unitlength=1.mm
	\begin{center}
		\begin{picture}(50,15)
		\put(12.5,4){$\kappa_{+} \qquad ,$}
		\put(0,0){\line(1,0){5}}
		\put(3,0){\vector(1,0){1}}
		\put(0,10){\line(1,0){5}}
		\put(3,10){\vector(-1,0){1}}
		\put(5,5){\oval(10, 10)[r]}
		\put(52.5,4){$\kappa_{-}$}
		\put(40,0){\line(1,0){5}}
		\put(43,0){\vector(-1,0){1}}
		\put(40,10){\line(1,0){5}}
		\put(43,10){\vector(1,0){1}}
		\put(45,5){\oval(10, 10)[r]}

		\end{picture}
		\caption{The non-trivial reflection matrix elements.}
		\label{Kmatrix}
	\end{center}
\end{figure}
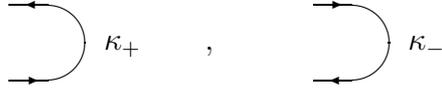

Nevertheless, the integrable structure can also be extended to tackle open boundary condition problems thanks to the Sklyanin construction \cite{SKYLIANIN}. In this context, the $R$-matrix continues describing the bulk dynamics and a new set of matrices, the so-called $K$-matrices, describes the interaction at the open ends. This is provided by reflection equation \cite{SKYLIANIN}, 
\eq
R_{12}(\lambda_1-\lambda_2)K_{1}(\lambda_1)R_{12}(\lambda_1+\lambda_2)K_{2}(\lambda_2) =K_{2}(\lambda_2) R_{12}(\lambda_1+\lambda_2) K_{1}(\lambda_1) R_{12}(\lambda_1-\lambda_2).
\label{refle}
\en
In the simplest case the $K$-matrix is diagonal
\eq
K(\lambda)=\left(\begin{array}{cc}
	\kappa_{+}(\lambda) & 0 \\
	0 & \kappa_{-}(\lambda)
\end{array}\right),
\en
and its matrix elements can be depicted as in Figure \ref{Kmatrix}.

The Sklyanin's monodromy matrix can be written as
\eq
U(\lambda)={\cal T}(\lambda)K(\lambda) \widetilde{\cal T}(\lambda)=\left(\begin{array}{cc}
	{\cal A}(\lambda) & {\cal B}(\lambda) \\
	{\cal C}(\lambda) & {\cal D}(\lambda) 
\end{array}\right),
\label{openT}
\en
where $\widetilde{\cal T}_j(\lambda)\propto \left[{\cal T}_j(-\lambda)\right]^{-1}$ is given by
\eq
\widetilde{\cal T}_j(\lambda)= R_{j 1}(\lambda+\mu_1)\cdots R_{j N}(\lambda+\mu_N)=\left(\begin{array}{cc}
	\widetilde{A}(\lambda) & \widetilde{B}(\lambda) \\
	\widetilde{C}(\lambda) & \widetilde{D}(\lambda) 
\end{array}\right).
\en

\begin{figure}[h]
	\unitlength=0.4mm
	\begin{center}
		\begin{picture}(80,130)(-30,-10)
		
		\multiput(-20,0)(0,20){6}{\line(1,0){70}}
		\multiput(0,-20)(20,0){3}{\line(0,1){140}}
		\multiput(50,10)(0,40){3}{\oval(20, 20)[r]}
		% arrow
		\multiput(-10,0)(0,20){6}{\vector(1,0){1}}
		\multiput(0,-12.5)(20,0){3}{\vector(0,-1){1}}
		\multiput(0,112.5)(20,0){3}{\vector(0,1){1}}

		\put(-35,-3){$\lambda_3$}
		\put(-40,17){$-\lambda_3$}
		\put(-35,37){$\lambda_2$}
		\put(-40,57){$-\lambda_2$}
		\put(-35,77){$\lambda_1$}
		\put(-40,97){$-\lambda_1$}
		
		\put(-5,125){$\mu_3$}
		\put(17,125){$\mu_2$}
		\put(37,125){$\mu_1$}
		
		\end{picture}
	\end{center}
	\caption{The partition function $Z_N$ for $N=3$ of the six-vertex model with reflecting end.}
	\label{ZDWRE}
\end{figure}
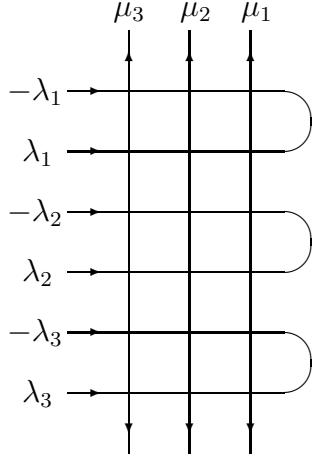

Now the product of Sklyanin monodromy matrix element ${\cal B}(\lambda)$ gives rise to another partition function for the six-vertex model due to Tsuchiya \cite{TSUCHIYA},
\eq
Z_N(\{\lambda\},\{\mu\})=\bra{\Downarrow}{\cal B}(\lambda_N)\cdots {\cal B}(\lambda_2) {\cal B}(\lambda_1)\ket{\Uparrow}.
\label{ZUturn}
\en
In this case, we still have domain wall like boundary on the vertical direction, but on the horizontal direction one has a reflecting end as illustrated in  Figure \ref{ZDWRE}.

The free energy of the six-vertex model with reflecting end was considered in the thermodynamic limit \cite{RIBEIRO2015a} and it was again shown to differ from the periodic boundary conditions. However, there is no result for correlation functions in the setting of reflecting end boundary condition.

\section{Boundary correlations}\label{boundcorr}

In this section we use the algebraic Bethe ansatz tools to the computation of boundary correlations for the six-vertex model with reflecting end. 

We fix the parametrization of Boltzmann weights as follows,
\begin{align}
a_{\pm}(\lambda_j,\mu_k)&=a(\lambda_j \pm \mu_k)=\sin(\lambda_j\pm\mu_k+2\eta),\nonumber \\
b_{\pm}(\lambda_j,\mu_k)&=b(\lambda_j \pm \mu_k)=\sin(\lambda_j\pm\mu_k),\label{par-2}\\
c_{\pm}(\lambda_j,\mu_k)&=c(\lambda_j \pm \mu_k)=\sin(2\eta),\nonumber
\end{align}
and also for the $K$-matrix elements,
\begin{align}
\kappa_{\pm}(\lambda)=\frac{\sin(\xi\pm\lambda)}{\sin(\xi)},
\end{align}
where $\xi$ is a boundary parameter. It is worth to mention that throughout this paper we use $c=c_{\pm}(\lambda,\mu)$.
Since the monodromy matrix $\mathcal{U}(\lambda_j)$ also satisfies the reflection equation,
\begin{align}
R_{12}(\lambda_1-\lambda_2)\mathcal{U}_1(\lambda_1)R_{12}(\lambda_1+\lambda_2)\mathcal{U}_2(\lambda_2)=\mathcal{U}_2(\lambda_2)R_{12}(\lambda_1+\lambda_2)\mathcal{U}_1(\lambda_1)R_{12}(\lambda_1-\lambda_2),
\label{refl-u}
\end{align}
we obtain the reflection algebra due to the procedure devised by Sklyanin \cite{SKYLIANIN}, whose main relations are given by,
\begin{align}
&[\mathcal{B}(\lambda_1),\mathcal{B}(\lambda_2)]=[\mathcal{C}(\lambda_1),\mathcal{C}(\lambda_2)]=0,\label{comm-bc}\\
&\mc{A}(\lambda_2)\mc{B}(\lambda_1)=f_1(\lambda_1,\lambda_2)\mc{B}(\lambda_1)\mc{A}(\lambda_2)+f_2(\lambda_1,\lambda_2)\mc{B}(\lambda_2)\mc{A}(\lambda_1)+f_3(\lambda_1,\lambda_2)\mc{B}(\lambda_2)\wt{\mc{D}}(\lambda_1),\label{comm-ac-1}\\
&\wt{\mc{D}}(\lambda_1)\mc{B}(\lambda_2)=g_1(\lambda_1,\lambda_2)\mc{B}(\lambda_2)\wt{\mc{D}}(\lambda_1)+g_2(\lambda_1,\lambda_2)\mc{B}(\lambda_1)\wt{\mc{D}}(\lambda_2)+g_3(\lambda_1,\lambda_2)\mc{B}(\lambda_1)\mc{A}(\lambda_2),\label{comm-dc-1}
\end{align}
where
\eq
\wt{\mathcal{D}}(\lambda)=\mathcal{D}(\lambda)-h(\lambda)\mathcal{A}(\lambda), \qquad \mbox{with} \qquad  h(\lambda)=\frac{c}{a(2\lambda)}, 
\en
and
\bear
f_1(\lambda_1,\lambda_2)&=&\frac{a_-(\lambda_1,\lambda_2)b_+(\lambda_1,\lambda_2)}{b_-(\lambda_1,\lambda_2)a_+(\lambda_1,\lambda_2)},\nonumber \\
f_2(\lambda_1,\lambda_2)&=&-\frac{cb_+(\lambda_1,\lambda_2)}{b_-(\lambda_1,\lambda_2)a_+(\lambda_1,\lambda_2)}-\frac{ch(\lambda_1)}{a_+(\lambda_1,\lambda_2)},\nonumber \\
f_3(\lambda_1,\lambda_2)&=&-\frac{c}{a_+(\lambda_1,\lambda_2)},\\
g_1(\lambda_1,\lambda_2)&=&\frac{a_-(\lambda_1,\lambda_2)a_+(\lambda_1,\lambda_2)}{b_-(\lambda_1,\lambda_2)b_+(\lambda_1,\lambda_2)}\left(1-\frac{c^2}{a_+^2(\lambda_1,\lambda_2)}\right),\nonumber \\ 
g_2(\lambda_1,\lambda_2)&=&-\frac{ca_+(\lambda_1,\lambda_2)}{b_-(\lambda_1,\lambda_2)b_+(\lambda_1,\lambda_2)}\left(1-\frac{c^2}{a_+^2(\lambda_1,\lambda_2)}\right)+\frac{ch(\lambda_1)}{a_+(\lambda_2,\lambda_1)},\nonumber \\
g_3(\lambda_1,\lambda_2)&=&h(\lambda_2)\left(g_2(\lambda_1,\lambda_2)-\frac{ch(\lambda_1)}{a_+(\lambda_2,\lambda_1)}\right)-h(\lambda_1)f_2(\lambda_2,\lambda_1)+\nonumber \\
&+&\frac{c a^2_-(\lambda_1,\lambda_2)}{a_+(\lambda_1,\lambda_2)b_-^2(\lambda_1,\lambda_2)}\left(1-\frac{c^2}{a_-^2(\lambda_1,\lambda_2)}\right). \nonumber
\ear

In addition, the action of the monodromy matrix operators over the reference state, which is taken as the ferromagnetic state, are given as
\bear
\mathcal{A}(\lambda)\ket{\Uparrow}&=&\kappa_+(\lambda)\alpha_+(\lambda)\alpha_-(\lambda)\ket{\Uparrow}=\beta(\lambda)\ket{\Uparrow}, \label{eigen-ac} \nonumber\\
\wt{\mathcal{D}}(\lambda)\ket{\Uparrow}&=&(\kappa_-(\lambda)-h(\lambda)\kappa_+(\lambda))\delta_+(\lambda)\delta_-(\lambda)\ket{\Uparrow}=\zeta(\lambda)\ket{\Uparrow}, \label{eigen-dc} \\ 
\mathcal{C}(\lambda)\ket{\Uparrow}&=&0 \nonumber,
\ear
where
\eq
\alpha_{\pm}(\lambda)=\prod_{k=1}^{N}a_{\pm}(\lambda,\mu_k), \qquad \delta_{\pm}(\lambda)=\prod_{k=1}^{N}b_{\pm}(\lambda,\mu_k).
\label{alphas}
\en

The action of the operators $\mc{A}(\lambda)$ and $\mc{D}(\lambda)$ over the off-shell Bethe state  $\mc{B}(\lambda_r)\ldots\mc{B}(\lambda_1)\ket{\Uparrow}$ can be obtained thanks to the relations (\ref{comm-bc})-(\ref{comm-dc-1}), yielding to
\begin{align}
&\mc{A}(\lambda)\prod_{j=1}^{r}\mc{B}(\lambda_j)\ket{\Uparrow}=\beta(\lambda)\prod_{j=1}^{r}f_1(\lambda_j,\lambda)\prod_{j=1}^{r}\mc{B}(\lambda_j)\ket{\Uparrow}+\nonumber \\
&+\sum_{i=1}^{r}\left(\beta(\lambda_i)f_2(\lambda_i,\lambda)\prod_{\substack{j=1\\j \ne i}}^{r}f_1(\lambda_j,\lambda_i)+\zeta(\lambda_i)f_3(\lambda_i,\lambda)\prod_{\substack{j=1\\j \ne i}}^{r}g_1(\lambda_i,\lambda_j)\right)\prod_{\substack{j=1\\j \ne i}}^{r}\mc{B}(\lambda_j)\ket{\Uparrow},
\label{AfB}\\
&\wt{\mc{D}}(\lambda)\prod_{j=1}^{r}\mc{B}(\lambda_j)\ket{\Uparrow}=\zeta(\lambda)\prod_{j=1}^{r}g_1(\lambda,\lambda_j)\prod_{j=1}^{r}\mc{B}(\lambda_j)\ket{\Uparrow}+\nonumber \\
&+\sum_{i=1}^{r}\left(\zeta(\lambda_i)g_2(\lambda,\lambda_i)\prod_{\substack{j=1\\j \ne i}}^{r}g_1(\lambda_i,\lambda_j)+\beta(\lambda_i)g_3(\lambda,\lambda_i)\prod_{\substack{j=1\\j \ne i}}^{r}f_1(\lambda_j,\lambda_i)\right)\prod_{\substack{j=1\\j \ne i}}^{r}\mc{B}(\lambda_j)\ket{\Uparrow}.
\label{DfB}
\end{align}

Now we have the main ingredients to introduce the boundary correlations for the six-vertex model with reflecting end boundary. We first start with two types of correlation functions describing the local state probabilities at the boundary, which were inspired by the ones defined for the domain wall boundary \cite{BOGOLIUBOV}. The first one describes a kind of boundary spontaneous polarization 
\eq
G_N^{(r)}=\frac{1}{Z_N}\bra{\Downarrow}\mc{B}(\lambda_N)\ldots\mc{B}(\lambda_{r+1})q_N\mc{B}(\lambda_r)\ldots\mc{B}(\lambda_1)\ket{\Uparrow},\label{g-refl-def}
\en 
while the second one describes the probability of having a $c$-type vertex at the intersection between the $N$-th column and one of the stripes of the $r$-th double row, which can be defined as
\eq
H_N^{(r)}=\frac{1}{Z_N}\bra{\Downarrow}\mc{B}(\lambda_N)\ldots\mc{B}(\lambda_{r+1})q_N\mc{B}(\lambda_r)p_N\mc{B}(\lambda_{r-1})\ldots\mc{B}(\lambda_1)\ket{\Uparrow},\label{h-refl-def}\\
\en
where $p_j=\frac{1}{2}(1+\sigma^z_j)$ and $q_j=\frac{1}{2}(1-\sigma^z_j)$ are the projectors on the spin-up and down states acting on the $j$-th column.

The above boundary correlations \eqref{g-refl-def}-\eqref{h-refl-def}, defined on a $2 N\times N$ lattice, can be expressed in terms of sums over the partition functions of $2(N-1)\times (N-1)$ rectangular sublattices. This is done by use of the commutation relations \eqref{comm-bc}-\eqref{comm-dc-1} and the decomposition of the monodromy in two parts, which is called the two-site model \cite{BOOK},
\begin{align}
\mc{T}_j(\lambda_j)=\mc{T}_{jN}(\lambda_j)	\mc{T}_{j1}(\lambda_j),&&\wt{\mc{T}}_j(\lambda_j)=\wt{\mc{T}}_{j1}(\lambda_j)	\wt{\mc{T}}_{jN}(\lambda_j),
\end{align}
where
\begin{align}
\mc{T}_{jN}(\lambda_j)=R_{jN}(\lambda_j-\mu_N),&&\mc{T}_{j1}(\lambda_j)=R_{jN-1}(\lambda_j-\mu_{N-1})\ldots R_{j1}(\lambda_j-\mu_1),\\
\wt{\mc{T}}_{jN}(\lambda_j)=R_{jN}(\lambda_j+\mu_N),&&\wt{\mc{T}}_{j1}(\lambda_j)=R_{j1}(\lambda_j+\mu_{1})\ldots R_{jN-1}(\lambda_j+\mu_{N-1}).
\end{align}
Again we represent the monodromy matrices $\mc{T}_{jk}$ and $\tilde{\mc{T}}_{jk}$ as
\begin{align}
\mc{T}_{jk}=\left(\begin{array}{cc}
A_{k} & B_{k} \\
C_{k} & D_{k}
\end{array}\right),&& \wt{\mc{T}}_{jk}=\left(\begin{array}{cc}
\wt{A}_{k} & \wt{B}_{k} \\
\wt{C}_{k} & \wt{D}_{k}
\end{array}\right),
\end{align}
where $k=1,N$. Therefore, we can decompose the Sklyanin's monodromy matrix as
\begin{align}
\mc{U}(\lambda_j)&=\mc{T}_{jN}(\lambda_j)[	\mc{T}_{j1}(\lambda_j)\mc{K}(\lambda_j)\wt{\mc{T}}_{j1}(\lambda_j)]\wt{\mc{T}}_{jN}(\lambda_j)\nonumber\\&=\mc{T}_{jN}(\lambda_j)\mc{U}_1(\lambda_j)\wt{\mc{T}}_{jN}(\lambda_j),		
\end{align}
where the matrix elements of $\mc{U}_1(\lambda_j)$, denoted as  $\mc{A}_1$, $\mc{B}_1$, $\mc{C}_1$ and $\mc{D}_1$ are operators acting on the vertical space except for the $N$-th column. This way, we can conveniently write the monodromy matrix element $\mc{B}$ as follows,
\begin{align}
\mc{B}(\lambda)=(A_N\mc{A}_1+B_N\mc{C}_1)\wt{B}_N+(A_N\mc{B}_1+B_N\mc{D}_1)\wt{D}_N,
\label{bcalzao}
\end{align}
where $B_N=\wt{B}_N=c\ \sigma_N^-$. Using (\ref{bcalzao}), we can reduce the problem of calculating the scalar products in \eqref{g-refl-def} and \eqref{h-refl-def} to the problem involving the operators $\mc{A}_1$, $\mc{B}_1$, $\mc{C}_1$ e $\mc{D}_1$. To see this, let us first compute the product $\mc{B}(\lambda_{r})\ldots\mc{B}(\lambda_1)\ket{\Uparrow}$. Substituting $B_N$ and $\wt{B}_N$ in \eqref{bcalzao}, we can bring $\mc{B}(\lambda)$ into the form
\begin{align}
	\mc{B}(\lambda)=c\mc{P}(\lambda){\sigma_N^-}+\mc{B}_1(\lambda)\mc{Q}_N(\lambda),
\end{align}
where
\begin{align}
	\mc{P}(\lambda)=A_N \mc{A}_1 + [\wt{D}_N+b_+(\lambda,\mu_N)-a_+(\lambda,\mu_N)]\mc{D}_1,&&\mc{Q}_N(\lambda)=A_N \wt{D}_N.
	\label{pq}
\end{align}
Note that the states $\ket{\uparrow}$, $\ket{\downarrow}$ are eigenvectors of $\mc{P}(\lambda)$ and $\mc{Q}_N(\lambda)$. Since $(\sigma_N^-)^2=0$, we have
\begin{align}
	\mc{B}(\lambda_{r})\ldots\mc{B}(\lambda_1)&\ket{\Uparrow}=\left[\prod_{j=1}^{r}a_-(\lambda_j,\mu_N)b_+(\lambda_j,\mu_N)\right]\mc{B}_1(\lambda_{r})\ldots\mc{B}_1(\lambda_1)\ket{\Uparrow}+\nonumber\\
	&+\sum_{i=1}^{r}\left[\prod_{j=i+1}^{r}\mc{B}_1(\lambda_j)\mc{Q}_N(\lambda_j)\right]c\mc{P}(\lambda_i)\sigma_N^-\left[\prod_{j=1}^{i-1}\mc{B}_1(\lambda_j)\mc{Q}_N(\lambda_j)\right]\ket{\Uparrow}.
	\label{prod-b}
\end{align}
Decomposing the ferromagnetic state as $\ket{\Uparrow}=\ket{\Uparrow}_1\otimes\ket{\uparrow}_N$, $\ket{\Uparrow}_1=\otimes_{k=1}^{N-1} \ket{\uparrow}_k$ and acting $\mc{P}$ and $\mc{Q}$ over $\ket{\uparrow}_N$, we can rewrite the sum in \eqref{prod-b} as
\begin{align}
	c&\sum_{i=1}^{r}\left[\prod_{j=1}^{i-1}a_-(\lambda_j,\mu_N)b_+(\lambda_j,\mu_N)\prod_{j=i+1}^{r}a_+(\lambda_j,\mu_N)b_-(\lambda_j,\mu_N)\right]\mc{B}_1(\lambda_{r})\ldots\mc{B}_1(\lambda_{i+1})\nonumber \\
	&\times[b_-(\lambda_i,\mu_N)\mc{A}_1(\lambda_i)+b_+(\lambda_i,\mu_N)\mc{D}_1(\lambda_i)]\mc{B}_1(\lambda_{i-1})\ldots\mc{B}_1(\lambda_1)\ket{\Uparrow}_1\otimes\ket{\downarrow}_N.
	\label{sum}
\end{align}
Since $p_N\ket{\downarrow}_N=0$, it is clear from \eqref{sum} that acting $p_N$ from the left on \eqref{prod-b} leads to 
\begin{align}
	p_N\mc{B}(\lambda_{r})\ldots\mc{B}(\lambda_1)\ket{\Uparrow}=\left[\prod_{j=1}^{r}a_-(\lambda_j,\mu_N)b_+(\lambda_j,\mu_N)\right]\mc{B}_1(\lambda_{r})\ldots\mc{B}_1(\lambda_1)\ket{\Uparrow}.
	\label{int-p}
\end{align}
Similarly,
\begin{align}
	\bra{\Downarrow}\mc{B}(\lambda_N)\ldots\mc{B}(\lambda_{r+1})q_N=\left[\prod_{j=r+1}^{N}a_+(\lambda_j,\mu_N)b_-(\lambda_j,\mu_N)\right]\bra{\Downarrow}\mc{B}_1(\lambda_N)\ldots\mc{B}_1(\lambda_{r+1}).
	\label{int-q}
\end{align}
Substituting \eqref{int-p} and \eqref{int-q} in \eqref{h-refl-def}, we arrive at
\begin{align}
H_N^{(r)}&=\frac{c}{Z_N}\prod_{j=1}^{r-1}a_-(\lambda_j,\mu_N)b_+(\lambda_j,\mu_N)\prod_{j=r+1}^{N}a_+(\lambda_j,\mu_N)b_-(\lambda_j,\mu_N) \nonumber \\
&\times \bra{\Downarrow}_1\mc{B}_1(\lambda_N)\ldots\mc{B}_1(\lambda_{r+1})\left[\bra{\downarrow}_N\mc{B}(\lambda_r)\ket{\uparrow}_N\right]\mc{B}_1(\lambda_{r-1})\ldots\mc{B}_1(\lambda_1)\ket{\Uparrow}_1 \nonumber \\
&=\frac{c}{Z_N}\prod_{j=1}^{r-1}a_-(\lambda_j,\mu_N)b_+(\lambda_j,\mu_N)\prod_{j=r+1}^{N}a_+(\lambda_j,\mu_N)b_-(\lambda_j,\mu_N)\nonumber \\
&\times \Big[ \bra{\Downarrow}_1\mc{B}_1(\lambda_N)\ldots\mc{B}_1(\lambda_{r+1})[(b_-(\lambda_r,\mu_N)+b_+(\lambda_r,\mu_N)h(\lambda_r))\mc{A}_1(\lambda_r)+ \nonumber \\
&+ b_+(\lambda_r,\mu_N)\wt{\mc{D}}_1(\lambda_r)]\mc{B}_1(\lambda_{r-1})\ldots\mc{B}_1(\lambda_1)\ket{\Uparrow}_1\Big].
\end{align}
Using \eqref{AfB} and \eqref{DfB}, we obtain the expression for the correlation function $H_N^{(r)}$,
\begin{align}
H_N^{(r)}&=\frac{c}{Z_N}\prod_{j=1}^{r-1}a_-(\lambda_j,\mu_N)b_+(\lambda_j,\mu_N)\prod_{j=r+1}^{N}a_+(\lambda_j,\mu_N)b_-(\lambda_j,\mu_N)\nonumber \\
&\times \sum_{i=1}^{r} S_H(\lambda_i) Z_{N-1}[\lambda_i;\mu_N],
\label{h-refl-rec}
\end{align}
where $S_H(\lambda_i)=S_{H,1}(\lambda_i)+S_{H,2}(\lambda_i)$, with
\begin{align}
S_{H,1}(\lambda_i)&=\left[\frac{(b_-(\lambda_r,\mu_N)+b_+(\lambda_r,\mu_N)h(\lambda_r))f_2(\lambda_i,\lambda_r)+b_+(\lambda_r,\mu_N)g_3(\lambda_r,\lambda_i)}{f_1(\lambda_r,\lambda_i)}\right]\nonumber \\
&\times \beta_1(\lambda_i)\prod_{\substack{j=1\\j\ne i}}^{r}f_1(\lambda_j,\lambda_i),\\
S_{H,2}(\lambda_i)&=\left[\frac{(b_-(\lambda_r,\mu_N)+b_+(\lambda_r,\mu_N)h(\lambda_r))f_3(\lambda_i,\lambda_r)+b_+(\lambda_r,\mu_N)g_2(\lambda_r,\lambda_i)}{g_1(\lambda_i,\lambda_r)}\right]\nonumber \\
&\times \zeta_1(\lambda_i)\prod_{\substack{j=1\\j\ne i}}^{r}g_1(\lambda_i,\lambda_j),
\end{align}
 and $Z_{N-1}[\lambda_i;\mu_N]$ is the partition function corresponding to a $2(N-1)\times (N-1)$ lattice, which is obtained from the $2N\times N$ lattice after excluding the column and double row associated to the parameters $\mu_N$ and $\lambda_i$, respectively,
\begin{align}
	Z_{N-1}[\lambda_i;\mu_N]&=\bra{\Downarrow}_1\mc{B}_1(\lambda_r)\ldots\mc{B}_1(\lambda_{i+1})\mc{B}_1({\lambda_{i-1}})\ldots\mc{B}_1(\lambda_1)\ket{\Uparrow}_1.
\end{align}

In order to obtain the correlation function $G_N^{(r)}$, we exploit the relation $G_N^{(r)}=H_N^{(1)}+\ldots+H_N^{(r)}$, 
\begin{align}
G_N^{(r)}&=\frac{c}{Z_N}\sum_{i=1}^{r}\prod_{j=1}^{i-1}a_-(\lambda_j,\mu_N)b_+(\lambda_j,\mu_N)\prod_{j=i+1}^{N}a_+(\lambda_j,\mu_N)b_-(\lambda_j,\mu_N) \nonumber \\
&\times\Big[\bra{\Downarrow}_1\mc{B}_1(\lambda_N)\ldots\mc{B}_1(\lambda_{i+1})[(b_-(\lambda_i,\mu_N)+b_+(\lambda_i,\mu_N)h(\lambda_i))\mc{A}_1(\lambda_i)+ \nonumber \\
&+ b_+(\lambda_i,\mu_N)\wt{\mc{D}}_1(\lambda_i)]\mc{B}_1(\lambda_{i-1})\ldots\mc{B}_1(\lambda_1)\ket{\Uparrow}_1\Big].
\label{g-sum}
\end{align}
This is done by noticing that the last term in the sum ($i=r$) is given by, 
\begin{align}
&	\prod_{j=1}^{r}a_-(\lambda_j,\mu_N)b_+(\lambda_j,\mu_N)\prod_{j=r+1}^{N}a_+(\lambda_j,\mu_N)b_-(\lambda_j,\mu_N) \nonumber \\
&\times \left[\frac{(b_-(\lambda_r,\mu_N)+b_+(\lambda_r,\mu_N)h(\lambda_r))}{a_-(\lambda_r,\mu_N)b_+(\lambda_r,\mu_N)}\beta_1(\lambda_r)\prod_{j=1}^{r-1}f_1(\lambda_j,\lambda_r)+\frac{\zeta_1(\lambda_r)}{a_-(\lambda_r,\mu_N)}  \nonumber \right.\\
&\left. \times  \prod_{j=1}^{r-1}g_1(\lambda_r,\lambda_j)\right]\bra{\Downarrow}_1\mc{B}_1(\lambda_N)\ldots\mc{B}_1(\lambda_{r+1})\mc{B}_1(\lambda_{r-1})\ldots\mc{B}_1(\lambda_1)\ket{\Uparrow}_1+(\Diamond),
\label{expl}
\end{align} 
where $\beta_1(\lambda)$ and $\zeta_1(\lambda)$ are eigenvalues of $\mc{A}_1(\lambda)$ and $\wt{\mc{D}}_1(\lambda)$, respectively, associated with the state $\ket{\Uparrow}_1$, and ``$(\Diamond)$'' includes terms depending on $\mc{B}_1(\lambda_r)$. The remaining part of the expression \eqref{expl} is the only one from \eqref{g-sum} that do not depend on $\mc{B}_1(\lambda_r)$.
Moreover, $G_N^{(r)}$ is a symmetric function of the variables $\lambda_1,\ldots,\lambda_r$, thanks to \eqref{g-refl-def} and \eqref{comm-bc}. Therefore, the contribution of the $k$-th term in the sum \eqref{g-sum} must be similar to the one written explicitly in \eqref{expl} (aside from the products before the brackets), with $r\rightarrow k$. Thus the final expression can be written as
\begin{align}
G_N^{(r)}&=\frac{c}{Z_N}\prod_{j=1}^{r}a_-(\lambda_j,\mu_N)b_+(\lambda_j,\mu_N)\prod_{j=r+1}^{N}a_+(\lambda_j,\mu_N)b_-(\lambda_j,\mu_N)\nonumber \\
&\times\sum_{i=1}^{r}S_G(\lambda_i)Z_{N-1}[\lambda_i;\mu_N],
\label{g-refl-rec}
\end{align}
where 
\begin{align}
S_G(\lambda_i)&= \frac{(b_-(\lambda_i,\mu_N)+b_+(\lambda_i,\mu_N)h(\lambda_i))}{a_-(\lambda_i,\mu_N)b_+(\lambda_i,\mu_N)}\beta_1(\lambda_i)\prod_{\substack{j=1\\j \ne i}}^{r}f_1(\lambda_j,\lambda_i)+\frac{\zeta_1(\lambda_i)}{a_-(\lambda_i,\mu_N)}\nonumber \\
&\times \prod_{\substack{j=1\\j \ne i}}^{r}g_1(\lambda_i,\lambda_j).
\end{align}

One can obtain a recursion formula for the partition function $Z_N$ from the expression \eqref{g-refl-rec}  for $r=N$,
\begin{align}
Z_N=c\prod_{j=1}^{N}a_-(\lambda_j,\mu_N)b_+(\lambda_j,\mu_N)\sum_{i=1}^{N}S_G(\lambda_i)Z_{N-1}[\lambda_i;\mu_N],
\label{zn-refl-rec}
\end{align}
by using the fact that $G_N^{(N)}=1$. This recursion relation is a generalization of the specialized relation obtained in \cite{TSUCHIYA}. The iteration of \eqref{zn-refl-rec} admits a determinant solution due to Tsuchiya \cite{TSUCHIYA} given as,
	\begin{align}
	Z_N&=\frac{\prod_{j,k=1}^{N}a_+(\lambda_j,\mu_k)a_-(\lambda_j,\mu_k)b_+(\lambda_j,\mu_k)b_-(\lambda_j,\mu_k)}{\prod_{k<j}^{N}a_+(\lambda_j,\lambda_k)b_-(\lambda_j,\lambda_k)\prod_{m<n}^{N}b_+(\mu_m,\mu_n)b_-(\mu_m,\mu_n)} \prod_{j=1}^{N}b(2\lambda_j)\kappa_{-}(\mu_j)\nonumber \\
	&\times\det\mathsf{M},
	\label{zn-refl-det}
	\end{align}
where the elements of the matrix $\ms{M}$ are given by
\begin{align}
\mathsf{M}_{jk}=\psi(\lambda_j,\mu_k),&&j,k=1,\ldots,N,&&\psi(\lambda,\mu)=\frac{c(\lambda,\mu)}{a_+(\lambda,\mu)a_-(\lambda,\mu)b_+(\lambda,\mu)b_-(\lambda,\mu)}.
\label{psi1}
\end{align}

Determinant expressions for the correlations $H_N^{(r)}$ and $G_N^{(r)}$ can be obtained from the recursion relations \eqref{h-refl-rec} and \eqref{g-refl-rec}, respectively, once we use the representation \eqref{zn-refl-det} for the partition function. In order to find such expressions, we first compute the ratio between $Z_N$ and $Z_{N-1}$, which reads
\begin{align}
\frac{Z_{N-1}[\lambda_i;\mu_N]}{Z_N}&=\frac{(-1)^{i-1}}{b(2\lambda_i)\kappa_-(\mu_N)}\frac{1}{\alpha_{1,-}(\lambda_i)\alpha_{1,-}(\lambda_i)\delta_{1,+}(\lambda_i)\delta_{1,-}(\lambda_i)} \prod_{j=1}^{N}\frac{\psi(\lambda_j,\mu_N)}{c} \nonumber \\
&\times\prod_{\substack{j=1\\j \ne i}}^{N}a_+(\lambda_j,\lambda_i)b_-(\lambda_j,\lambda_i) \prod_{k=1}^{N-1}b_+(\mu_k,\mu_N)b_-(\mu_k,\mu_N)\frac{\det\mathsf{M}_{[i;N]}}{\det\mathsf{M}},
\label{ratio}
\end{align}
where $\ms{M}_{[j;k]}$, $j,k=1,\ldots,N$ denotes the $(N-1)\times (N-1)$ matrix obtained from $\ms{M}$ by removing its $j$-th row and $k$-th column, and
\begin{align}
\alpha_{1,\pm}(\lambda)=\prod_{k=1}^{N-1}a_{\pm}(\lambda,\mu_k),&&\delta_{1,\pm}(\lambda)=\prod_{k=1}^{N-1}b_{\pm}(\lambda,\mu_k).
\label{alphas1}
\end{align}
Then, substituting \eqref{ratio} in \eqref{h-refl-rec} and canceling the common factors, we have that
\begin{align}
	H_N^{(r)}&=\frac{\psi(\lambda_r,\mu_N)}{\kappa_{-}(\mu_N)\det\mathsf{M}}\frac{(-1)^{N-1}\prod_{k=1}^{N-1}b_+(\mu_k,\mu_N)b_-(\mu_k,\mu_N)}{\prod_{j=1}^{r-1}a_+(\lambda_j,\mu_N)b_-(\lambda_j,\mu_N)\prod_{j=r+1}^{N}a_-(\lambda_j,\mu_N)b_+(\lambda_j,\mu_N)} \nonumber \\
	&\times \sum_{i=1}^{r}(-1)^{i+N}u_r(\lambda_i)\det\mathsf{M}_{[i;N]},
	\label{h-sum}
\end{align}
where $u_r(\lambda)=u_{r,1}(\lambda)+u_{r,2}(\lambda)$, with
\begin{align}
	u_{r,1}(\lambda)&=\left[\frac{(b_-(\lambda_r,\mu_N)+b_+(\lambda_r,\mu_N)h(\lambda_r))f_2(\lambda,\lambda_r)+b_+(\lambda_r,\mu_N)g_3(\lambda_r,\lambda)}{f_1(\lambda_r,\lambda)}\right]\nonumber \\
	&\times \frac{\kappa_+(\lambda)}{cb^2(2\lambda)\delta_{1,+}(\lambda)\delta_{1,-}(\lambda)}\prod_{j=r+1}^{N}a_+(\lambda_j,\lambda)b_-(\lambda_j,\lambda)\prod_{j=1}^{r}a_-(\lambda_j,\lambda)b_+(\lambda_j,\lambda), 
\end{align}
\begin{align}
	u_{r,2}(\lambda)&=\left[\frac{(b_-(\lambda_r,\mu_N)+b_+(\lambda_r,\mu_N)h(\lambda_r))f_3(\lambda,\lambda_r)+b_+(\lambda_r,\mu_N)g_2(\lambda_r,\lambda)}{g_1(\lambda,\lambda_r)}\right]\nonumber \\
	&\times \frac{(\kappa_-(\lambda)-\kappa_+(\lambda)h(\lambda))}{c b(2\lambda)(b(2\lambda)-2\Delta a(2\lambda))\alpha_{1,+}(\lambda)\alpha_{1,-}(\lambda)}\prod_{j=r+1}^{N}a_+(\lambda_j,\lambda)b_-(\lambda_j,\lambda)  \nonumber\\
	&\times \prod_{j=1}^{r}a_-(\lambda,\lambda_j)[b_+(\lambda,\lambda_j)-2\Delta a_+(\lambda,\lambda_j)],
	\label{us}
\end{align}
where $a(\lambda)=a_{\pm}(\lambda,0)$, $b(\lambda)=b_{\pm}(\lambda,0)$. Note that $u_r(\lambda_i)=0$ for $i=r+1,\ldots,N$. Therefore, we can extend the sum in \eqref{h-sum} to $N$ and interpret it as the determinant of some $N\times N$ matrix developed by its last column. Hence, $H_N^{(r)}$ can be rewritten as
\begin{align}
	H_N^{(r)}&=\frac{\psi(\lambda_r,\mu_N)}{\kappa_{-}(\mu_N)}\frac{(-1)^{N+1}\prod_{k=1}^{N-1}b_+(\mu_k,\mu_N)b_-(\mu_k,\mu_N)}{\prod_{j=1}^{r-1}a_+(\lambda_j,\mu_N)b_-(\lambda_j,\mu_N)\prod_{j=r+1}^{N}a_-(\lambda_j,\mu_N)b_+(\lambda_j,\mu_N)}\frac{\det\mathsf{H}}{\det\mathsf{M}},
	\label{h-refl-det}
\end{align}
where $\mathsf{H}$ is the matrix whose elements are
\begin{align}
	\mathsf{H}_{jk}=\begin{cases}
	\mathsf{M}_{jk},& k\ne N,\\
	u_r(\lambda_j),& k=N.
	\end{cases}
\end{align}

Following the same procedure, after substituting \eqref{ratio} in \eqref{g-refl-rec} we are left with
\begin{align}
	G_N^{(r)}&=\frac{c}{\kappa_{-}(\mu_N)\det\mathsf{M}}\frac{(-1)^{N+1}\prod_{k=1}^{N-1}b_+(\mu_k,\mu_N)b_-(\mu_k,\mu_N)}{\prod_{j=1}^{r}a_+(\lambda_j,\mu_N)b_-(\lambda_j,\mu_N)\prod_{j=r+1}^{N}a_-(\lambda_j,\mu_N)b_+(\lambda_j,\mu_N)} \nonumber \\
	&\times \sum_{i=1}^{r}(-1)^{i+N}t_r(\lambda_i)\det\mathsf{M}_{[i;N]},
	\label{g-sum-1}
\end{align}
where
\begin{align}
	t_r(\lambda)=\frac{t_{r,1}(\lambda)+t_{r,2}(\lambda)}{a_-(\lambda,\mu_N)b_+(\lambda,\mu_N)},
\end{align}
\begin{align}
t_{r,1}(\lambda)&=\frac{\kappa_+(\lambda)}{cb^2(2\lambda)}\frac{[b_-(\lambda,\mu_N)+b_+(\lambda,\mu_N)h(\lambda)]}{\delta_{1,+}(\lambda)\delta_{1,-}(\lambda)}\prod_{j=r+1}^{N}a_+(\lambda_j,\lambda)b_-(\lambda_j,\lambda) \nonumber \\
&\times \prod_{j=1}^{r}a_-(\lambda_j,\lambda)	b_+(\lambda_j,\lambda), \\
t_{r,2}(\lambda)&=\frac{b_+(\lambda,\mu_N)}{cb(2\lambda)(b(2\lambda)-2\Delta a(2\lambda))}\frac{[\kappa_-(\lambda)-\kappa_+(\lambda)h(\lambda)]}{\alpha_{1,+}(\lambda)\alpha_{1,-}(\lambda)}\prod_{j=r+1}^{N}a_+(\lambda_j,\lambda)b_-(\lambda_j,\lambda) \nonumber \\
&\times \prod_{j=1}^{r}a_-(\lambda,\lambda_j)[b_+(\lambda,\lambda_j) -2\Delta a_+(\lambda,\lambda_j)].
\label{ts}
\end{align}
It is worth to emphasize that $\lambda_1,\ldots,\lambda_N$, $\mu_1,\ldots,\mu_N$ are to be regarded as parameters in \eqref{us} and \eqref{ts}, $\lambda$ being the only variable of these functions. Moreover, similarly to $u_r(\lambda_i)$ we also have $t_r(\lambda_i)=0$ for $i=r+1,\ldots,N$, which allows for the expression \eqref{g-refl-det} to be cast in the form
\begin{align}
	G_N^{(r)}&=\frac{c}{\kappa_{-}(\mu_N)}\frac{(-1)^{N+1}\prod_{k=1}^{N-1}b_+(\mu_k,\mu_N)b_-(\mu_k,\mu_N)}{\prod_{j=1}^{r}a_+(\lambda_j,\mu_N)b_-(\lambda_j,\mu_N)\prod_{j=r+1}^{N}a_-(\lambda_j,\mu_N)b_+(\lambda_j,\mu_N)}\frac{\det\mathsf{G}}{\det\mathsf{M}},
	\label{g-refl-det}
\end{align}
where
\begin{align}
\mathsf{G}_{jk}=\begin{cases}
\mathsf{M}_{jk},& k\ne N,\\
t_r(\lambda_j),& k=N.
\end{cases}
\end{align}

\subsection{Homogeneous limit}
We are now interested in the homogeneous limit  $\lambda_1,\ldots,\lambda_N\rightarrow\lambda$, $\mu_1,\ldots,\mu_N\rightarrow\mu$ of the boundary correlations $H_N^{(r)}$ and $G_N^{(r)}$. This can be done along the lines of \cite{KOREPIN1992}. Let
\begin{align}
\lambda_j=\lambda+\xi_j,&&j=1,\ldots,N.
\label{lam}
\end{align}
Starting with \eqref{h-refl-det}, the term 
$\det{\ms{M}}$
is transformed into
\begin{align}
\lim_{\substack{\xi_1,\ldots,\xi_N\rightarrow 0\\\mu_1,\ldots,\mu_N\rightarrow \mu}}\det\ms{M}=\prod_{j=0}^{N-1}\frac{(\mu_{j+1}-\mu)^j}{j!}\frac{\xi_{j+1}^j}{j!}\det\overline{\ms{M}},&&\overline{\ms{M}}_{jk}=\del_{\lambda}^{j-1}\del_{\mu}^{k-1}\psi(\lambda,\mu).
\label{hom-m}
\end{align}

Now, consider the identity
\begin{align}
\exp(\xi\del_{\epsilon})v(\lambda+\epsilon)\Big{|}_{\epsilon=0}=v(\lambda+\xi),
\label{id}
\end{align}
for some analytical function $v(\lambda)$. Then, applying \eqref{id} to \eqref{h-refl-det} together with standard properties of determinants, $\det\mathsf{H}$ can be written as
\begin{align}
	\det\mathsf{H}=\begin{vmatrix}
	\psi(\lambda_1,\mu_1) &  \cdots & \psi(\lambda_1,\mu_{N-1}) & \exp(\xi_1\del_{\epsilon}) \\
	\psi(\lambda_2,\mu_1) &  \cdots & \psi(\lambda_2,\mu_{N-1}) & \exp(\xi_2\del_{\epsilon}) \\
	& \vdots & & \\
	\psi(\lambda_N,\mu_1) &  \cdots & \psi(\lambda_N,\mu_{N-1}) & \exp(\xi_N\del_{\epsilon})
	\end{vmatrix}u_r(\lambda+\epsilon)\Bigg{|}_{\epsilon=0}.
	\label{g-inter}
\end{align}
Then, the homogeneous limit of the determinant in \eqref{g-inter} is taken in the same way as $\det\mathsf{M}$, which leads to  
\begin{align}
	\lim_{\substack{\xi_1,\ldots,\xi_N\rightarrow 0\\\mu_1,\ldots,\mu_N\rightarrow \mu}}\det{\ms{H}}&=\prod_{j=0}^{N-2}\frac{(\mu_{j+1}-\mu)^j}{j!}\prod_{j=0}^{N-1}\frac{\xi_{j+1}^j}{j!}\det\ms{U}(\del_{\epsilon})u_r(\lambda+\epsilon)\Bigg{|}_{\epsilon=0},
	\label{h-det-hom}
\end{align}
where the elements of $\ms{U}(\del_{\epsilon})$ are
\begin{align}
	\mathsf{U}_{jk}(\del_{\epsilon})=\begin{cases}
	\del_{\lambda}^{j-1}\del_{\mu}^{k-1}\psi(\lambda,\mu),& k\ne N,\\
	\del_{\epsilon}^{j-1},&k=N.
	\end{cases}
\end{align}

Finally, for the prefactor,
\begin{align}
	\lim_{\substack{\lambda_1,\ldots,\lambda_N\rightarrow \lambda\\\mu_1,\ldots,\mu_N\rightarrow \mu}} &\frac{\psi(\lambda_r,\mu_N)}{\kappa_{-}(\mu_N)}\frac{\prod_{k=1}^{N-1}b_+(\mu_k,\mu_N)b_-(\mu_k,\mu_N)}{\prod_{j=1}^{r-1}a_+(\lambda_j,\mu_N)b_-(\lambda_j,\mu_N)\prod_{j=r+1}^{N}a_-(\lambda_j,\mu_N)b_+(\lambda_j,\mu_N)}\nonumber \\
	&=\frac{\psi(\lambda,\mu)}{\kappa_{-}(\mu)}\frac{[b(2\mu)]^{N-1}}{[a_+(\lambda,\mu)b_-(\lambda,\mu)]^{r-1}[a_-(\lambda,\mu)b_+(\lambda,\mu)]^{N-r}}(\mu-\mu_{N})^{N-1},
	\label{h-pref-hom}
\end{align}
Substituting \eqref{hom-m}-\eqref{h-pref-hom}  into \eqref{h-refl-det}, it follows that the homogeneous limit of $H_N^{(r)}$ is given by
\begin{align}
	H_N^{(r)}&=\frac{(N-1)!}{\det\overline{\ms{M}}}\frac{\psi(\lambda,\mu)}{\kappa_{-}(\mu)}\frac{[b(2\mu)]^{N-1}}{[a_+(\lambda,\mu)b_-(\lambda,\mu)]^{r-1}[a_-(\lambda,\mu)b_+(\lambda,\mu)]^{N-r}}\nonumber \\
	&\times\det\mathsf{U}(\del_{\epsilon})u_r(\lambda+\epsilon)\Bigg{|}_{\epsilon=0}.
	\label{hom-h}
\end{align}

Due to the similarities between expressions \eqref{h-refl-det} and \eqref{g-refl-det}, the homogeneous limit of $G_N^{(r)}$ goes along the same lines of the limit for the $H_N^{(r)}$ discussed above. For this reason, we simply present the final result, which reads
\begin{align}
G_N^{(r)}&=\frac{(N-1)!}{\det\overline{\ms{M}}}\frac{c}{\kappa_{-}(\mu)}\frac{[b(2\mu)]^{N-1}}{[a_+(\lambda,\mu)b_-(\lambda,\mu)]^{r}[a_-(\lambda,\mu)b_+(\lambda,\mu)]^{N-r}}\nonumber\\
&\times \det\mathsf{U}(\del_{\epsilon})t_r(\lambda+\epsilon)\Bigg{|}_{\epsilon=0}.
\label{hom-g}
\end{align}

\section{Emptiness formation probability}\label{EFP}

Finally, we would like to introduce another kind of correlation function, the so-called emptiness formation probability, which was introduced in the context of vertex model for domain wall boundary \cite{COLOMO}. In the case of reflecting end boundary,  correlation of this kind describes the probability of having arrows pointing down on the first $s$ vertical edges (counting from the left) between the $r$-th and $(r+1)$-th double rows, which can be defined as
\begin{align}
F_N^{(r,s)}=\frac{1}{Z_N}\bra{\Downarrow}\mc{B}(\lambda_N)\ldots\mc{B}(\lambda_{r+1})q_N \ldots q_{N-s+1}\mc{B}(\lambda_r)\ldots\mc{B}(\lambda_1)\ket{\Uparrow}.
\label{f-refl-def}
\end{align}
Note that $F_N^{(r,1)}=G_N^{(r)}$. Once again, we use the two-site model decomposition of the monodromy matrix in order to obtain a recursion relation for $F_N^{(r,s)}$. The substitution of  \eqref{bcalzao} in \eqref{f-refl-def} leads to the products \eqref{prod-b} and \eqref{int-q}, except this time the sum \eqref{sum} does not vanish. Thus we have that,
\begin{align}
	&F_N^{(r,s)}=\frac{c}{Z_N}\prod_{j=r+1}^{N}a_+(\lambda_j,\mu_N)b_-(\lambda_j,\mu_N)\bra{\Downarrow}_1\mc{B}_1(\lambda_N)\ldots\mc{B}_1(\lambda_{r+1})q_{N-1}\ldots q_{N-s+1} \nonumber\\
	&\times \sum_{i=1}^{r}\left[\prod_{j=1}^{i-1}a_-(\lambda_j,\mu_N)b_+(\lambda_j,\mu_N) \prod_{j=i+1}^{r}a_+(\lambda_j,\mu_N)b_-(\lambda_j,\mu_N)\right]\mc{B}_1(\lambda_r)\ldots\mc{B}_1(\lambda_{i+1})\nonumber \\
	&\times [(b_-(\lambda_i,\mu_N)+b_+(\lambda_i,\mu_N)h(\lambda_i))\mc{A}_1(\lambda_i)+b_+(\lambda_i,\mu_N)\wt{\mc{D}}_1(\lambda_i)]\mc{B}_1(\lambda_{i-1})\ldots \mc{B}_1(\lambda_1)\ket{\Uparrow}_1.
	\label{f-int}
\end{align}
Note that the sum in the expression \eqref{f-int} is almost equal as the one in \eqref{g-sum}. Indeed, the last term ($i=r$) in the sum \eqref{f-int} gives a similar expression to \eqref{expl}, except for the projector operators $q_{N-1},\ldots,q_{N-s+1}$. From the definition \eqref{f-refl-def}, it follows that the same arguments we used before to rearrange the sum \eqref{g-sum} apply here. Therefore, $F_N^{(r,s)}$ can be brought into the form
\begin{align}
F_N^{(r,s)}&=\frac{c}{Z_N}\prod_{j=1}^{r}a_-(\lambda_j,\mu_N)b_+(\lambda_j,\mu_N)\prod_{j=r+1}^{N}a_+(\lambda_j,\mu_N)b_-(\lambda_j,\mu_N) \nonumber \\
&\times \sum_{i=1}^{r}S_G(\lambda_i)Z_{N-1}[\lambda_i;\mu_N]F_{N-1}^{(r-1,s-1)}[\lambda_i;\mu_N],
\label{f-refl-rec}
\end{align}
where
\begin{align}
	F_{N-1}^{(r-1,s-1)}[\lambda_i;\mu_N]&=\frac{1}{Z_{N-1}[\lambda_i;\mu_N]}\bra{\Downarrow}_1\mc{B}_1(\lambda_N)\ldots\mc{B}_1(\lambda_{r+1})q_{N-1}\ldots q_{N-s+1}\nonumber \\
	&\times\mc{B}_1({\lambda_r})\ldots\mc{B}_1(\lambda_{i+1})\mc{B}_1(\lambda_{i-1})\ldots\mc{B}_1(\lambda_1)\ket{\Uparrow}_1.
\end{align}

Analogously to the other boundary correlations, we provide a determinant representation for the emptiness formation probability. In what follows, we obtain a explicit formula for $F_N^{(r,s)}$ by iteration of the recurrence relation.

We recall that $F_N^{(r,1)}=G_N^{(r)}$ for the case $s=1$ is simply taken from \eqref{g-sum-1}. For the case $s=2$, we use the recursion relation \eqref{f-refl-rec} together with the previous result with $r\rightarrow r-1,\ N\rightarrow N-1$ allowing us to obtain $F_{N-1}^{(r-1,1)}[\lambda_i;\mu_N]$. In order to eliminate from \eqref{g-sum-1} the terms that depend on $\lambda_i$ or $\mu_N$, the prefactor of the sum must be modified accordingly,
\begin{align}
(-1)^{N+1}\rightarrow (-1)^N,&& \kappa_-(\mu_N)\rightarrow \kappa_-(\mu_{N-1}),&&\det\ms{M}\rightarrow \det\ms{M}_{[i;N]},
\label{subs1}
\end{align}
\begin{align}	
&\prod_{k=1}^{N-1}b_+(\mu_k,\mu_N)b_-(\mu_k,\mu_N)\rightarrow\prod_{k=1}^{N-2}b_+(\mu_k,\mu_{N-1})b_-(\mu_k,\mu_{N-1}),\nonumber \\
&\prod_{j=r+1}^{N}a_-(\lambda_j,\mu_N)b_+(\lambda_j,\mu_N)\rightarrow \prod_{j=r+1}^{N}a_-(\lambda_j,\mu_{N-1})b_+(\lambda_j,\mu_{N-1}),\\
&\prod_{j=1}^{r}a_+(\lambda_j,\mu_N)b_-(\lambda_j,\mu_N)\rightarrow\frac{1}{a_+(\lambda_i,\mu_{N-1})b_-(\lambda_i,\mu_{N-1})} \prod_{j=1}^{r}a_+(\lambda_j,\mu_{N-1})b_-(\lambda_j,\mu_{N-1}), \nonumber
\end{align}
whereas the sum should be replaced by
\begin{align}
&\sum_{j=1}^{N}(-1)^{j+N}\left[\frac{t_{r,1}(\lambda_j)+t_{r,2}(\lambda_j)}{a_-(\lambda_j,\mu_N)b_+(\lambda_j,\mu_N)}\right]\det\ms{M}_{[j;N]}\rightarrow \label{subs2} \\ &\sum_{\substack{j=1\\j\ne i}}^{N}(-1)^{j+N+\theta(i,j)}\left[\frac{\chi_1(\lambda_i,\lambda_j,\mu_N,\mu_{N-1})t_{r,1}(\lambda_j)+\chi_2(\lambda_i,\lambda_j,\mu_N,\mu_{N-1})t_{r,2}(\lambda_j)}{a_-(\lambda_j,\mu_{N-1})b_+(\lambda_j,\mu_{N-1})}\right]\nonumber \\
&\times\det\ms{M}_{[i,j;N-1,N]},
\nonumber
\end{align}
with
\begin{align}
	\theta(i,j)=\begin{cases}
	1,&i>j, \\
	0,&i<j,
	\end{cases}
\end{align}
and
\begin{align}
\chi_1(\lambda_i,\lambda_j,\mu_N,\mu_{N-1})&=\left[\frac{b_-(\lambda_j,\mu_{N-1})+b_+(\lambda_j,\mu_{N-1})h(\lambda_j)}{b_-(\lambda_j,\mu_{N})+b_+(\lambda_j,\mu_{N})h(\lambda_j)}\right]\frac{b_+(\lambda_j,\mu_{N-1})b_-(\lambda_j,\mu_{N-1})}{a_-(\lambda_i,\lambda_j)b_+(\lambda_i,\lambda_j)},\\
\chi_2(\lambda_i,\lambda_j,\mu_N,\mu_{N-1})&=\frac{b_+(\lambda_j,\mu_{N-1})}{b_+(\lambda_j,\mu_{N})}\frac{a_+(\lambda_j,\mu_{N-1})a_-(\lambda_j,\mu_{N-1})}{a_-(\lambda_j,\lambda_i)[b_+(\lambda_j,\lambda_i) -2\Delta a_+(\lambda_j,\lambda_i)]}.
\end{align}
Using \eqref{subs1}-\eqref{subs2} to obtain $F_{N-1}^{(r-1,1)}$ and replacing it into the recursion formula \eqref{f-refl-rec} leave us with the following expression 
\begin{align}
&F_N^{(r,2)}=\prod_{k=N-1}^{N}\left[\frac{(-1)^{k+1} c}{\kappa_-(\mu_k)}\frac{\prod_{m=1}^{k-1}b_+(\mu_m,\mu_k)b_-(\mu_m,\mu_k)}{\prod_{j=1}^{r}a_+(\lambda_j,\mu_k)b_-(\lambda_j,\mu_k)\prod_{j=r+1}^{N}a_-(\lambda_j,\mu_k)b_+(\lambda_j,\mu_k)}\right] \nonumber \\
&\times \frac{1}{\det\ms{M}} \sum_{i=1}^{N}\sum_{\substack{j=1\\j\ne i}}^{N}(-1)^{(i+N)+(j+N)+\theta(i,j)}a_+(\lambda_i,\mu_{N-1})b_-(\lambda_i,\mu_{N-1}) \det\ms{M}_{[i,j;N-1,N]} \nonumber \\
&\times \frac{[t_{r,1}(\lambda_i)+t_{r,2}(\lambda_i)][{\chi_1(\lambda_i,\lambda_j,\mu_N,\mu_{N-1})t_{r,1}(\lambda_j)+\chi_2(\lambda_i,\lambda_j,\mu_N,\mu_{N-1})t_{r,2}(\lambda_j)}]}{a_-(\lambda_j,\mu_{N-1})b_+(\lambda_j,\mu_{N-1})a_-(\lambda_i,\mu_{N})b_+(\lambda_i,\mu_{N})}.
\label{f-2}
\end{align}

Repeating the same steps as above, we can use \eqref{f-2} to obtain $F_N^{(r,3)}$, which gives
\begin{align}
	&F_N^{(r,3)}=\prod_{k=N-2}^{N}\left[\frac{(-1)^{k+1} c}{\kappa_-(\mu_k)}\frac{\prod_{m=1}^{k-1}b_+(\mu_m,\mu_k)b_-(\mu_m,\mu_k)}{\prod_{j=1}^{r}a_+(\lambda_j,\mu_k)b_-(\lambda_j,\mu_k)\prod_{j=r+1}^{N}a_-(\lambda_j,\mu_k)b_+(\lambda_j,\mu_k)}\right] \nonumber \\
	&\times \frac{1}{\det\ms{M}} \sum_{j_1=1}^{N}\sum_{\substack{j_2=1\\j_2 \ne j_1}}^{N}\sum_{\substack{j_3=1\\j_3 \ne j_1,j_2}}^{N}(-1)^{\sum_{i=1}^{3}(j_i+N)+\sum_{ m<n}^{3}\theta(j_m,j_n)} \det\ms{M}_{[j_1,j_2,j_3;N-2,N-1,N]} \nonumber \\
	&\times  \prod_{i=1}^{3}\left[\frac{\sum_{m=1,2}t_{r,m}(\lambda_{j_i})\prod_{k=1}^{i-1}\chi_{m}(\lambda_{j_k},\lambda_{j_i},\mu_{N-k+1},\mu_{N-k})}{a_-(\lambda_{j_i},\mu_{N-i+1})b_+(\lambda_{j_i},\mu_{N-i+1})}\right]\nonumber\\
	&\times \prod_{i<k}^{3}a_+(\lambda_{j_i},\mu_{N-k+1})b_-(\lambda_{j_i},\mu_{N-k+1}).
	\label{f-3}
\end{align}

By inspection of the expressions \eqref{g-refl-det}, \eqref{f-2} and \eqref{f-3}, we have the following formula for generic $s$,
\begin{align}
&F_N^{(r,s)}=\prod_{k=N-s+1}^{N}\left[\frac{(-1)^{k+1}c}{\kappa_-(\mu_k)}\frac{\prod_{m=1}^{k-1}b_+(\mu_m,\mu_k)b_-(\mu_m,\mu_k)}{\prod_{j=1}^{r}a_+(\lambda_j,\mu_k)b_-(\lambda_j,\mu_k)\prod_{j=r+1}^{N}a_-(\lambda_j,\mu_k)b_+(\lambda_j,\mu_k)}\right] \nonumber \\
&\times \frac{1}{\det\ms{M}} \sum_{j_1=1}^{N}\sum_{\substack{j_2=1\\j_2 \ne j_1}}^{N}\cdots\sum_{\substack{j_s=1\\j_s \ne j_{q< s}}}^{N}(-1)^{\sum_{i=1}^{s}(j_i+N)+\sum_{m<n}^{s}\theta(j_m,j_n)} \det\ms{M}_{[j_1,\ldots,j_s;N-s+1,\ldots,N]}\nonumber \\
&\times  \prod_{i=1}^{s}\left[\frac{\sum_{m=1,2}t_{r,m}(\lambda_{j_i})\prod_{k=1}^{i-1}\chi_{m}(\lambda_{j_k},\lambda_{j_i},\mu_{N-k+1},\mu_{N-k})}{a_-(\lambda_{j_i},\mu_{N-i+1})b_+(\lambda_{j_i},\mu_{N-i+1})}\right]\nonumber\\
&\times \prod_{i<k}^{s}a_+(\lambda_{j_i},\mu_{N-k+1})b_-(\lambda_{j_i},\mu_{N-k+1}).
\label{f-refl-sum}
\end{align}
It is worth noting that the general expression \eqref{f-refl-sum} satisfies the recursion relation \eqref{f-refl-rec}.

The sum in \eqref{f-refl-sum} can be rewritten as the determinant of a $N\times N$ matrix  by using the identity \eqref{id} (and \eqref{lam}). Thus we have that,
\begin{align}
&F_N^{(r,s)}=\prod_{k=N-s+1}^{N}\left[\frac{(-1)^{k+1}c}{\kappa_-(\mu_k)}\frac{\prod_{m=1}^{k-1}b_+(\mu_m,\mu_k)b_-(\mu_m,\mu_k)}{\prod_{j=1}^{r}a_+(\lambda_j,\mu_k)b_-(\lambda_j,\mu_k)\prod_{j=r+1}^{N}a_-(\lambda_j,\mu_k)b_+(\lambda_j,\mu_k)}\right] \nonumber \\
&\times \frac{\det\ms{V}(\del_{\epsilon_1},\ldots,\del_{\epsilon_s})}{\det\ms{M}}\left\{\prod_{i<k}^{s}a_+(\lambda+\epsilon_i,\mu_{N-k+1})b_-(\lambda+\epsilon_i,\mu_{N-k+1})  \nonumber \right. \\
&\left. \times 	
\prod_{i=1}^{s}\left[\frac{\sum_{m=1,2}t_{r,m}(\lambda+\epsilon_i)\prod_{k=1}^{i-1}\chi_{m}(\lambda+\epsilon_k,\lambda+\epsilon_i,\mu_{N-k+1},\mu_{N-k})}{a_-(\lambda+\epsilon_i,\mu_{N-i+1})b_+(\lambda+\epsilon_i,\mu_{N-i+1})}\right]\right\}_{\epsilon_1=\ldots=\epsilon_{s}=0},
\label{f-refl-det}
\end{align}
with the elements of $\ms{V}(\del_{\epsilon_1},\ldots,\del_{\epsilon_s})$ given by
\begin{align}
\ms{V}_{jk}(\del_{\epsilon_1},\ldots,\del_{\epsilon_s})=\begin{cases}
	\psi(\lambda+\xi_j,\mu_k),&k\le N-s,\\
	\exp(\xi_j\del_{\epsilon_{N-k+1}}),& k>N-s.
\end{cases}
\end{align}

\subsection{Homogeneous limit}
In order to compute the homogeneous limit of the emptiness formation probability $F_N^{(r,s)}$, we follow the same reasoning as in the previous section. For the determinant $\det\mathsf{V}(\del_{\epsilon_1},\ldots,\del_{\epsilon_s})$, we obtain
\begin{align}
\lim_{\substack{\xi_1,\ldots,\xi_N\rightarrow 0\\\mu_1,\ldots,\mu_N\rightarrow \mu}}	\det\ms{V}(\del_{\epsilon_1},\ldots,\del_{\epsilon_s})=\prod_{j=0}^{N-s-1}\frac{(\mu_{j+1}-\mu)^j}{j!}\prod_{j=0}^{N-1}\frac{\xi_{j+1}^j}{j!}\det\overline{\ms{V}},
\label{hom-v}
\end{align}
where the entries of  $\overline{\ms{V}}(\del_{\epsilon_1},\ldots,\del_{\epsilon_s})$ are
\begin{align}
	\overline{\ms{V}}_{jk}(\del_{\epsilon_1},\ldots,\del_{\epsilon_s})=\begin{cases}
	\del_{\lambda}^{j-1}\del_{\mu}^{k-1}\psi(\lambda,\mu),& k\le N-s \\
	\del_{\epsilon_{N-k+1}}^{j-1},& k> N-s.
	\end{cases}
\end{align}
On the other hand, the homogeneous limit of the remaining factors is straightforward:
\begin{align}
&\lim_{\substack{\xi_1,\ldots,\xi_N\rightarrow 0\\\mu_1,\ldots,\mu_N\rightarrow \mu}}\prod_{k=N-s+1}^{N}\left[\frac{(-1)^{k+1}c}{\kappa_-(\mu_k)}\frac{\prod_{m=1}^{k-1}b_+(\mu_m,\mu_k)b_-(\mu_m,\mu_k)}{\prod_{j=1}^{r}a_+(\lambda_j,\mu_k)b_-(\lambda_j,\mu_k)\prod_{j=r+1}^{N}a_-(\lambda_j,\mu_k)b_+(\lambda_j,\mu_k)}\right]\nonumber \\
&=\prod_{k=N-s+1}^{N} \left[\frac{(-1)^{k+1}c}{\kappa_-(\mu)}\frac{1}{[a_-(\lambda,\mu)b_+(\lambda,\mu)]^{N-r}[a_+(\lambda,\mu)b_-(\lambda,\mu)]^{r}}\right]\prod_{j=0}^{s-1}[b(2\mu)]^{N-s+j}\nonumber \\
&\times \prod_{j=N-s}^{N-1}(\mu-\mu_{j+1})^j,
\label{hom-prod}
\end{align}
\begin{align}
&\lim_{\mu_1,\ldots,\mu_N\rightarrow \mu}\prod_{i<k}^{s}a_+(\lambda+\epsilon_i,\mu_{N-k+1})b_-(\lambda+\epsilon_i,\mu_{N-k+1}) \nonumber \\
&\times 
\prod_{i=1}^{s}\left[\frac{\sum_{m=1,2}t_{r,m}(\lambda+\epsilon_i)\prod_{k=1}^{i-1}\chi_{m}(\lambda+\epsilon_k,\lambda+\epsilon_i,\mu_{N-k+1},\mu_{N-k})}{a_-(\lambda+\epsilon_i,\mu_{N-i+1})b_+(\lambda+\epsilon_i,\mu_{N-i+1})}\right]\nonumber \\
&=\prod_{i=1}^{s-1} [a_+(\lambda+\epsilon_i,\mu)b_-(\lambda+\epsilon_i,\mu)]^{s-i} \prod_{i=1}^{s}\left[\frac{\sum_{m=1,2}t_{r,m}(\lambda+\epsilon_i)\prod_{k=1}^{i-1}\chi_{m}(\lambda+\epsilon_k,\lambda+\epsilon_i,\mu)}{a_-(\lambda+\epsilon_i,\mu)b_+(\lambda+\epsilon_i,\mu)} \right].
\label{hom-r}
\end{align}
Substituting \eqref{hom-v}-\eqref{hom-r} and \eqref{hom-m} in \eqref{f-refl-det}, we obtain the emptiness formation probability in the homogeneous limit,
\begin{align}
F_N^{(r,s)}&=\frac{1}{\det\overline{\ms{M}}}\prod_{k=N-s+1}^{N}\left[\frac{c(k-1)!}{\kappa_-(\mu)}\frac{[b(2\mu)]^{k-1}}{[a_-(\lambda,\mu)b_+(\lambda,\mu)]^{N-r}[a_+(\lambda,\mu)b_-(\lambda,\mu)]^{r}}\right] \nonumber \\
&\times\det\overline{\ms{V}}(\del_{\epsilon_1},\ldots,\del_{\epsilon_s})\left\{\prod_{i=1}^{s-1} [a_+(\lambda+\epsilon_i,\mu)b_-(\lambda+\epsilon_i,\mu)]^{s-i} \nonumber \right. \\
&\left.\times \prod_{i=1}^{s}\left[\frac{\sum_{m=1,2}t_{r,m}(\lambda+\epsilon_i)\prod_{k=1}^{i-1}\chi_{m}(\lambda+\epsilon_k,\lambda+\epsilon_i,\mu)}{a_-(\lambda+\epsilon_i,\mu)b_+(\lambda+\epsilon_i,\mu)} \right]
\right\}_{\epsilon_1=\ldots=\epsilon_{s}=0}.
\label{f-refl-hom}
\end{align}

\subsection{Biorthogonal polynomials representation}
Another equivalent determinant representation for emptiness formation probability can be obtained by means of biorthogonal polynomials. In this representation, the function $F_N^{(r,s)}$ is given in terms of the determinant of a $s\times s$ matrix, in contrast to the $N\times N$ determinant appearing in \eqref{f-refl-hom}. In order to derive such representation, we must first lay out some facts regarding biorthogonal polynomials \cite{BERTOLA}.

Let $\{P_n(x)\}$, $\{Q_m(y)\}$, $n,m=0,1,\ldots$ be two polynomial sequences of one real variable, satisfying
\begin{align}
	\int_{-\infty}^{\infty}\int_{-\infty}^{\infty}P_n(x) Q_m(y)w(x,y)\dd x\dd y=\delta_{nm}J_n,
\end{align}
where $w(x,y)$ is a weight function, $\delta_{nm}$ is the Kronecker delta and $J_n$, $n=0,1,\ldots$ are constants. Then $P_n(x)$ and $Q_m(y)$ are said to be biorthogonals. Also, consider the moments $m_{j,k}$, defined as
\begin{align}
	m_{j,k}=\int_{-\infty}^{\infty}\int_{-\infty}^{\infty}x^j y^k w(x,y)\dd x \dd y.
\end{align}
Assuming $P_n$ and $Q_m$ are monic polynomials, the following relations are valid: 
\begin{align}
	\begin{vmatrix}
	m_{0,0} & m_{0,1} & \ldots & m_{0,n-1} \\
	m_{1,0} & m_{1,1} & \ldots & m_{1,n-1} \\
	& & \vdots & \\
	m_{n-1,0} & m_{n-1,1} & \ldots & m_{n-1,n-1}
	\end{vmatrix}=J_0 J_1\ldots J_{n-1},
	\label{biorth-1}
\end{align} 
\begin{align}
\begin{vmatrix}
m_{0,0} &  \ldots & m_{0,n-2} & 1 \\
m_{1,0} &  \ldots & m_{1,n-2} & x \\
&  \vdots & & \vdots \\
m_{n-1,0} & \ldots & m_{n-1,n-2} & x^{n-1}
\end{vmatrix}=J_0 J_1\ldots J_{n-2}P_{n-1}(x),
\label{biorth-2}
\end{align}  
\begin{align}
	\begin{vmatrix}
	m_{0,0}  & \cdots & m_{0,n-l-1} & 1 & \cdots & 1 \\
	m_{1,0}  & \cdots & m_{1,n-l-1} & x_1 & \cdots & x_l \\
	&  \vdots & &  & \vdots & \\
	m_{n-1,0}  & \cdots & m_{n-1,n-l-1} & x_1^{n-1} & \cdots & x_l^{n-1} 
	\end{vmatrix}=J_0 J_1 \ldots J_{n-l-1} \begin{vmatrix}
	P_{n-l}(x_1) & \cdots & P_{n-l}(x_l) \\
	& \vdots & \\
	P_{n-1}(x_1) & \cdots & P_{n-1}(x_l)
	\end{vmatrix}.
\label{biorth-3}	
\end{align} 
 
Identifying the moments as $m_{j,k}=\del_{\lambda}^{j-1}\del_{\mu}^{k-1}[c(\lambda,\mu)\psi(\lambda,\mu)]$ where the $\psi(\lambda,\mu)$ is defined in \eqref{psi1} and the weight function
\begin{align}
w(x,y)=\frac{1}{2}e^{(\lambda+\eta)x+\mu y}\Phi\left(\frac{x-y}{2}\right)\Phi\left(\frac{x+y}{2}\right),&&\Phi(x)=e^{-\pi x/2}\frac{\sinh(\eta x)}{\sinh(\pi x/2)},
\end{align}
after comparing $\det\overline{\ms{M}}$ and $\det\overline{\ms{V}}$ with relations \eqref{biorth-1}-\eqref{biorth-3}, we have
\begin{align}
	\det\overline{\ms{M}}&=\left(\frac{1}{c}\right)^N J_0\ldots J_{N-1},\\
	\det\overline{\ms{V}}(\del_{\epsilon_1},\ldots,\del_{\epsilon_s})&=\left(\frac{1}{c}\right)^{N-s} J_0 J_1 \ldots J_{N-s-1} \begin{vmatrix}
	P_{N-s}(\del_{\epsilon_s}) & \cdots & P_{N-s}(\del_{\epsilon_1}) \\
	& \vdots & \\
	P_{N-1}(\del_{\epsilon_s}) & \cdots & P_{N-1}(\del_{\epsilon_1})
	\end{vmatrix}.
	\label{dets}
\end{align}
Substituting \eqref{dets} in \eqref{f-refl-hom}, we obtain the biorthogonal polynomial representation for emptiness formation probability,
\begin{align}
F_N^{(r,s)}&=\prod_{k=N-s+1}^{N}\left[\frac{c^2}{\kappa_-(\mu)}\frac{[b(2\mu)]^{k-1}}{[a_-(\lambda,\mu)b_+(\lambda,\mu)]^{N-r}[a_+(\lambda,\mu)b_-(\lambda,\mu)]^{r}}\right]\nonumber\\
&\times\begin{vmatrix}
V_{N-s}(\del_{\epsilon_s}) & \cdots & V_{N-s}(\del_{\epsilon_1}) \\
& \vdots & \\
V_{N-1}(\del_{\epsilon_s}) & \cdots & V_{N-1}(\del_{\epsilon_1})
\end{vmatrix}\left\{\prod_{i=1}^{s-1} [a_+(\lambda+\epsilon_i,\mu)b_-(\lambda+\epsilon_i,\mu)]^{s-i} \nonumber \right. \\
&\left.\times \prod_{i=1}^{s}\left[\frac{\sum_{m=1,2}t_{r,m}(\lambda+\epsilon_i)\prod_{k=1}^{i-1}\chi_{m}(\lambda+\epsilon_k,\lambda+\epsilon_i,\mu)}{a_-(\lambda+\epsilon_i,\mu)b_+(\lambda+\epsilon_i,\mu)} \right]
\right\}_{\epsilon_1=\ldots=\epsilon_{s}=0},
\label{f-refl-poly}
\end{align}
where $V_n(x)=(n!/J_m)P_n(x)$.

As a special case, let $s=1$ in \eqref{f-refl-poly}. Then, it follows that $G_N^{(r)}$ in the homogeneous limit can also be written as
\begin{align}
	G_N^{(r)}&=\frac{c^2}{\kappa_-(\mu)}\frac{[b(2\mu)]^{N-1}}{[a_-(\lambda,\mu)b_+(\lambda,\mu)]^{N-r}[a_+(\lambda,\mu)b_-(\lambda,\mu)]^{r}}V_{N-1}(\del_{\epsilon})t_r(\lambda+\epsilon)\Bigg{|}_{\epsilon=0}.
\end{align}
In addition, since $H_N^{(r)}=G_N^{(r)}-G_N^{(r-1)}$,
\begin{align}
	&H_N^{(r)}=\frac{c^2}{\kappa_-(\mu)}\frac{[b(2\mu)]^{N-1}}{[a_-(\lambda,\mu)b_+(\lambda,\mu)]^{N-r}[a_+(\lambda,\mu)b_-(\lambda,\mu)]^{r}}\nonumber\\
	&\times V_{N-1}(\del_{\epsilon})\left\{\frac{1}{a_-(\lambda+\epsilon,\mu)b_+(\lambda+\epsilon,\mu)}\left[t_{r,1}(\lambda+\epsilon)\left(1-\frac{1}{f_1(\lambda,\mu)f_1(\lambda,\lambda+\epsilon)}\right) \nonumber \right.\right. \\
	&\left.\left.  +t_{r,2}(\lambda+\epsilon)\left(1+\frac{1}{f_1(\lambda,\mu)g_1(\lambda+\epsilon,\lambda)}\right)\right]\right\}_{\epsilon=0}.
\end{align}

In order to derive the arctic curves for the six-vertex model with reflecting end boundary along the same lines as \cite{PRONKO}, the next step would be obtain a multiple integral representation for the emptiness formation probability in the homogeneous limit and apply the saddle-point method, from which the parametric equations for a portion of the curve are expected to follow. Although we have an expression for this correlation in terms of biorthogonal polynomials, certain analytical properties which would allow us to transform this representation into an integral one are still missing. 

\section{Conclusion}\label{conclusion}

In this paper we computed boundary correlation functions of the six-vertex model with domain wall and reflecting end boundary. This comprises the explicit calculation of the boundary spontaneous polarization and the emptiness formation probability. In order to do that we exploit the  Sklyanin's reflection algebra and the Tsuchiya determinant representation for the partition function with reflecting end boundary. The homogeneous limit was taken and therefore the correlations were expressed in terms of determinant of $N\times N$ matrices. In case of the emptiness formation probability, we have further simplified the final expression in terms of the determinant of a $s\times s$ matrix, thanks to the use of biorthogonal polynomials properties. 

We hope that the our results could be useful in the study of the arctic curves for the six-vertex model with reflecting end boundary.  The existence of such spatial phase separation was recently studied numerically \cite{LYBERG} and it would be interesting to have an analytical description for the arctic curves.

\section*{Acknowledgments}

The authors thank the S\~ao Paulo Research Foundation (FAPESP) for financial support through the grants 2017/22363-9 and 2017/16535-1. 
G.A.P. Ribeiro thanks the Simons Center for Geometry and Physics of Stony Brook University and the organizers of the scientific program "Exactly Solvable Models of Quantum Field Theory and Statistical Mechanics" 
for hospitality and support during part of this work. He also thanks for the hospitality of the Bergische Wuppertal University where part of this work was written.

\end{document}